%% file: ms.tex
\documentclass[preprint2]{aastex6}

\usepackage{amsmath}

\shortauthors{Sheehan et al.}
\shorttitle{GY 91 Disk Gaps}

\begin{document}

\title{Multiple Gaps in the Disk of the Class I Protostar GY 91}
\author{Patrick D. Sheehan\altaffilmark{1} and Josh A. Eisner\altaffilmark{1}}
\affil{\altaffilmark{1}Steward Observatory, University of Arizona, 933 N. Cherry Avenue, Tucson, AZ, 85721}

%\email{psheehan@email.arizona.edu}

\begin{abstract}
We present the highest spatial resolution ALMA observations to date of the Class I protostar GY 91 in the $\rho$ Ophiuchus L1688 molecular cloud complex. Our 870 $\mu$m and 3 mm dust continuum maps show that the GY 91 disk has a radius of $\sim$80 AU, and an inclination of $\sim$40$^{\circ}$, but most interestingly that the disk has three dark lanes located at 10 AU, 40 AU, and 70 AU. We model these features assuming they are gaps in the disk surface density profile and find that their widths are 7 AU, 30 AU, and 10 AU. These gaps bear a striking resemblance to the gaps seen in the HL Tau disk, suggesting that there may be Saturn-mass planets hiding in the disk. To constrain the relative ages of GY 91 and HL Tau, we also model the disk and envelope of HL Tau and find that they are of similar ages, although GY 91 may be younger. Although snow lines and magnetic dead zones can also produce dark lanes, if planets are indeed carving these gaps then Saturn-mass planets must form within the first $\sim$0.5 Myr of the lifetime of protoplanetary disks.
\end{abstract}

\section{Introduction}

Planets form in protoplanetary disks. When observed at high resolution with ALMA, a number of these disks show interesting patterns in their millimeter emission profiles, in some cases including series of bright and dark rings \citep{Brogan2015,Dong2015,Andrews2016,Isella2016,Loomis2017,Fedele2017,Fedele2017b,Dipierro2018}. Some of these disks may even be quite young ($\lesssim1$ Myr) \citep{Brogan2015,Dong2015,Fedele2017b,Dipierro2018}. There are a number of explanations for such features, including chemical processes that alter dust opacities and sticking/fracturing processes near snow lines \citep[e.g.][]{Ros2013,Zhang2015,Banzatti2015} as well as vortices created at the edges of magnetic dead zones \citep[e.g.][]{Simon2014,Flock2015}, but the most exciting possibility is that these features are tracing gaps opened in disks by forming planets \citep[e.g.][]{Dong2015}.

GY 91 is a M4 protostar \citep{Doppmann2005} in the L1688 region of the $\rho$ Ophiuchus molecular cloud, located at a distance of 137 pc \citep{OrtizLeon2017}. GY 91's broadband spectral energy distribution (SED) rises sharply in the infrared and appears to peak at far-infrared wavelengths, although it has not been detected between 35 $\mu$m and 870 $\mu$m. The infrared spectral index ($\alpha_{IR} = 0.45$) and bolometric temperature ($T_{bol} = 370$ K), as well as its association with a 1.1 mm core, classify GY 91 as a Class I protostar \citep[e.g.][]{Enoch2008,McClure2010,Dunham2015}. This indicates that the protostar is surrounded by a protoplanetary disk still embedded in its natal envelope of collapsing cloud material, and is young \citep[$\lesssim0.5$ Myr;][]{Evans2009}. The Spitzer IRS spectrum of the source shows both silicate and ice absorption features, which are also commonly associated with embedded protostars \citep[e.g.][]{Watson2004}.

A few studies that consider alternate classification schemes have suggested that GY 91 may not be embedded. \citet{McClure2010} find that the 5--12 $\mu$m spectral index is within the range found for disks with foreground extinction ($n_{5-12} = -0.25$). However, their measured value is also on the border between disks with foreground extinction and disks with envelopes (of $n_{5-12} = -0.2$), and the extinction corrected spectral index ($\alpha_{IR}' = 0.31$) and bolometric temperature ($T_{bol}' = 470$ K) still qualify the source as a Class I protostar  \citep{Dunham2015}.  \citet{vanKempen2009} also found HCO$^+$ emission towards GY 91 that was bright enough to be above the cutoff for an embedded source, but that emission seems to be associated with a patch of cloud that peaks 30" away from the source.

\floattable
\input{table1.tex}

Here we present new ALMA data, which when combined with the observed SED, show that GY 91 is indeed a Class I protostar with a circumstellar disk embedded in an envelope. Our 3 mm and 870 $\mu$m images also reveal the presence of three narrow dark rings in its disk that resemble those seen in HL Tau and a handful of other disks \citep{Brogan2015,Dong2015,Andrews2016,Isella2016,Loomis2017,Fedele2017,Fedele2017b,Dipierro2018}. We compare the circumstellar structure of GY 91 to HL Tau, and argue that GY 91 is the youngest source in which disk gaps have been detected.  If caused by planets, these features provide evidence for giant planet formation within ~0.5 Myr.

\section{Observations \& Data Reduction}

\subsection{ALMA}

GY 91 was observed with ALMA Band 3 (100 GHz/3 mm) in three tracks from 31 October 2015 to 17 April 2016, with baselines ranging from 14 m -- 15.3 km. All four basebands were tuned for continuum observations centered at 90.5, 92.5, 102.5, 104.5 GHz, each with 128 15.625 MHz channels for 2 GHz of continuum bandwidth per baseband. In all the observations had 8 GHz of total continuum bandwidth. We also observed GY 91 with ALMA Band 7 (345 GHz/870 $\mu$m) on 19 May 2016 and 11 September 2016, with baselines ranging from 15 -- 3140 m. Two of four basebands were configured for continuum observations centered at 343 GHz and 356.25 GHz, with a total of 4 GHz of continuum bandwidth. The remaining basebands were devoted to spectral line observations, although nothing was detected. We list details of the observations in Table \ref{table:alma_obs}.

The data were reduced in the standard way with the \texttt{CASA} pipeline and the calibrators listed in Table \ref{table:alma_obs}. After calibrating, we imaged the data by Fourier transforming the visibilities with the \texttt{CLEAN} routine. After our initial imaging, we found that we could improve the sensitivity of the 345 GHz image by self-calibrating. We ran four iterations of phase-only self-calibration on the compact configuration track and a single iteration of phase-only self-calibration on the extended configuration track. This improved the rms in an image produced with natural weighting (i.e. a robust parameter of 2) from 0.36 mJy /beam to 0.27 mJy/beam. We were unable to improve the 100 GHz image by self-calibration.

\begin{figure*}
\centering
\figurenum{1}
\includegraphics[width=7in]{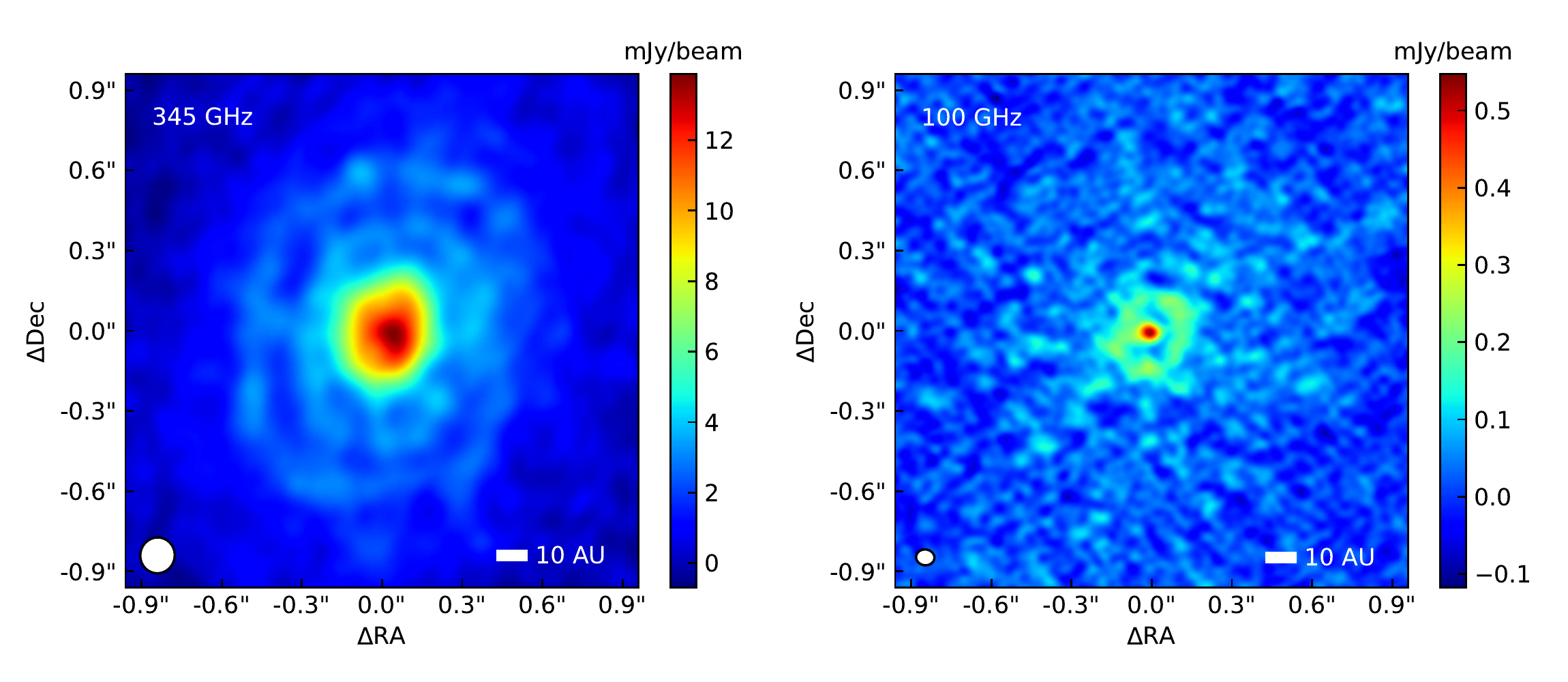}
\caption{Our ALMA 345 GHz ({\it left}) and 100 GHz ({\it right}) maps of the GY 91 protoplanetary disk. Two dark lanes are readily apparent in the 345 GHz map, while a third dark lane is also apparent in inner regions of the disk at 100 GHz because of the higher resolution of our 100 GHz maps.}
\label{fig:alma_data}
\end{figure*}

Our final images were produced using Briggs weighting with a robust parameter of 0.5, which provides a good balance between sensitivity and resolution, to weight the visibilities for both datasets. The 3 mm image has a beam of size 0.06" by 0.05" with a P.A. of 81.9$^{\circ}$ and an rms of 36 $\mu$Jy/beam. The 870 $\mu$m image nas a beam size of 0.134" by 0.129" with a P.A. of -9.4$^{\circ}$ and an rms of 0.31 mJy/beam. We show the images in Figure \ref{fig:alma_data}.

\subsection{SED from the Literature}

We compile a broadband spectral energy distribution (SED) for GY 91 from a thorough literature search. The data includes {\it Spitzer} IRAC and MIPS photometry as well as fluxes from the literature at a range of wavelengths \citep{Wilking1983,Lada1984,Greene1992,Andre1994,Strom1995,Barsony1997,Johnstone2000,Allen2002,Natta2006,Stanke2006,AlvesdeOliveira2008,Jorgensen2008,Padgett2008,Wilking2008,Evans2009,Gutermuth2009,Barsony2012}. When modeling the SED, as we discuss below, we assume a constant 10\% uncertainty on any photometry from the literature to account for any flux calibration uncertainties between the measurements.

In addition to the broadband photometry, we also download the Spitzer IRS spectrum of GY 91 from the CASSIS database \citep{Lebouteiller2011,Lebouteiller2015}. Rather than consider the entire SED, which can be computationally prohibitive for the radiative transfer calculations described below, we sample the IRS spectrum at 25 points ranging from 5 to 35 $\mu$m. We also assume a 10\% uncertainty on these fluxes, like we do for the broadband photometry.

\begin{figure}[b]
\centering
\figurenum{2}
\includegraphics[width=3.3in]{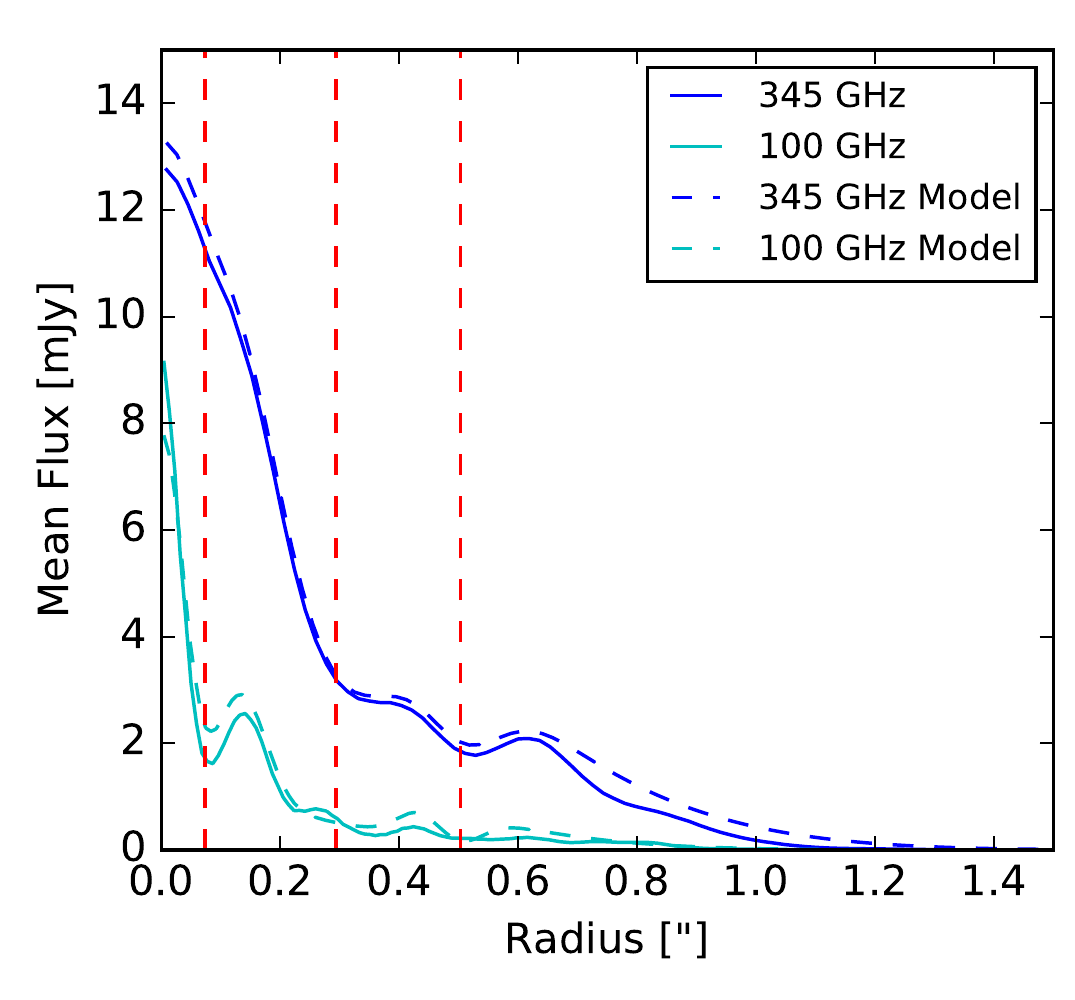}
\caption{The one dimensional, azimuthally averaged, de-projected radial brightness profile of the GY 91 disk at 345 GHz and 100 GHz, with the locations of the dark lanes marked by vertical dashed lines. These gaps are readily seen in the brightness profile. We also show the azimuthally averaged brightness profile of our gapped disk+envelope model (see Figure \ref{fig:disk_ulrich}, Table 2) at each wavelength.}
\label{fig:radial_profile}
\end{figure}

\section{Results}

We show our 3mm and 870 $\mu$m images of GY 91 in Figure \ref{fig:alma_data}. The 870 $\mu$m image has a much higher signal-to-noise ratio, and it is fairly easy to identify, by-eye, two concentric dark lanes that appear in the disk. The 3 mm image also reveals a third dark lane in the inner regions of the disk that is not visible at 870 $\mu$m because of the factor of two poorer resolution. As the disk is massive (see below), high optical depth in the inner disk could also help to hide the inner gap at 870 $\mu$m. To better illustrate the presence of these features, in Figure \ref{fig:radial_profile} we show a one dimensional brightness profile for both the 870 $\mu$m and 3 mm images, averaged in ellipses defined by the position angle and inclination of the disk to be constant radius bins. The outer two dark lanes show up clearly in the 870 $\mu$m radial profile, while the inner lane shows up clearly in the 3 mm profile. Moreover, there appears to be a break in the 870 $\mu$m profile at the location of the inner dark lane, and there appears to be a dip in the 3 mm brightness profile that is consistent with the location of the middle dark lane, despite the noisiness of the 3 mm image that prevents it from being detected by eye.

\begin{figure*}
\centering
\figurenum{3}
\includegraphics[width=7in]{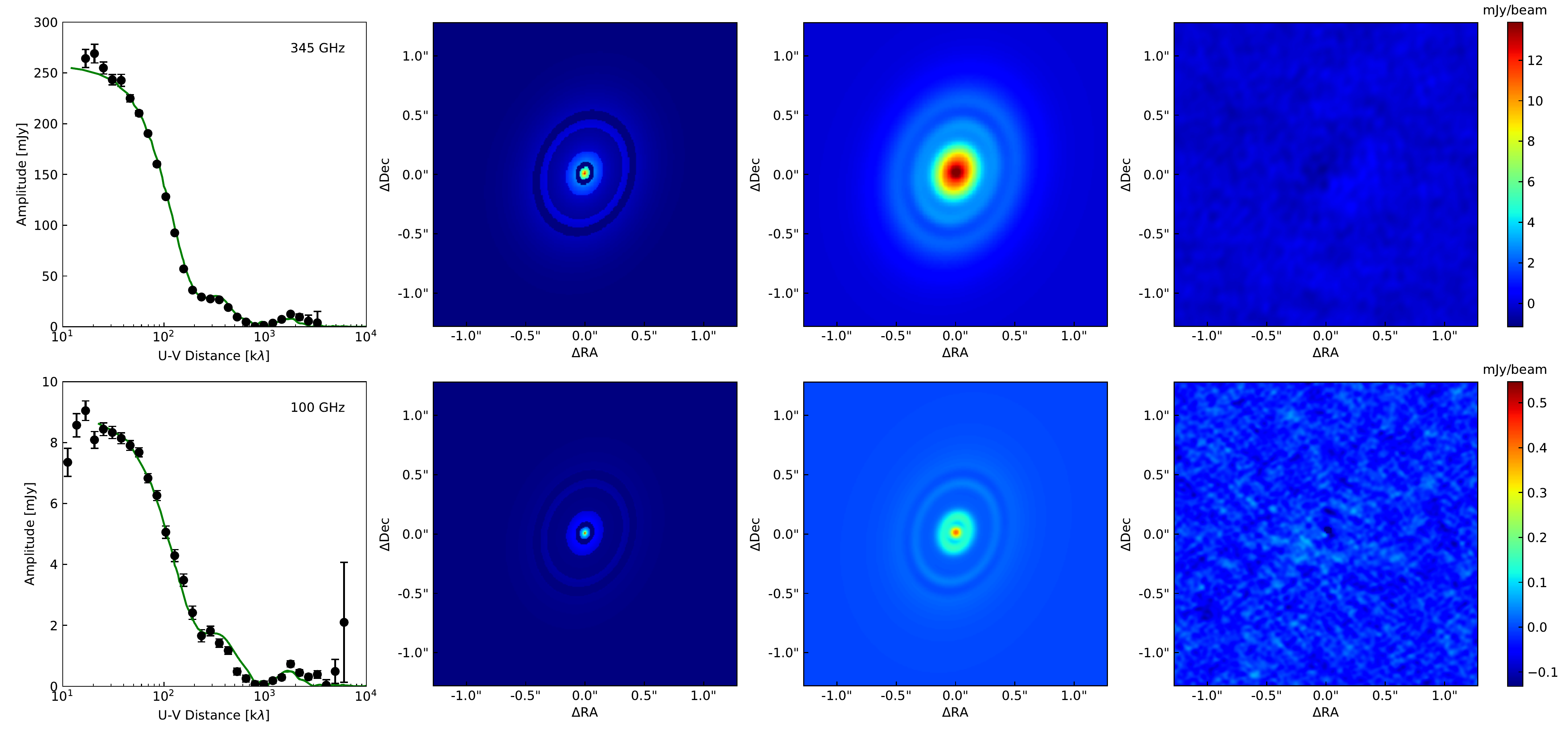}
\caption{The best fit simple geometrical model for GY 91 compared with the data. The model assumes the disk is flat, with a surface density described by Equation 1 and $M_{disk} = 0.36$ M$_{\odot}$, $r_c = 71$ AU, $\gamma = 0.3$, $i = 39^{\circ}$, and $p.a. = -19^{\circ}$. The model uses a power-law temperature distribution with $T = 46 (R / 1 \, \text{AU})^{-0.4}$. We use the power-law millimeter opacity function described in \citet{Beckwith1990}, $\kappa({\nu}) = 0.1 \, (\nu / 1000 \, \text{GHz})^\beta$ cm$^2$ g$^{-1}$ with $\beta = 1.8$. Our model includes three gaps with the following parameters: $R_{gap,1} = 10.4$ AU, $w_{gap,1} = 5.9$ AU, $\delta_{gap,1} \approx 0$, $R_{gap,2} = 40.3$ AU, $w_{gap,2} = 27.5$ AU, $\delta_{gap,2} = 0.15$, $R_{gap,3} = 68.9$ AU, $w_{gap,3} = 10.7$ AU, and $\delta_{gap,3} \approx 0$. Our modeling indicates that the first and third gaps are deep, however as the data is noisy and not high enough resolution to well resolve the gaps, the actual depths are quite uncertain. As the disk is quite massive, it is somewhat optically thick at $870$ $\mu$m. We show the one dimensional, azimuthally averaged, visibility amplitudes on the left, the raw model images on the center-left, the model images convolved with the beam in the center-right column, and the residuals on the right. The colorbar shows the color scale for the center-right and right-most panels. The peak residuals are $1.7\sigma$ at 345 GHz and $3.5\sigma$ at 100 GHz.}
\label{fig:geometrical_model}
\end{figure*}

In order to study these features in greater detail, we fit a model to the data to determine disk properties such as radius, position angle, and inclination, as well as the locations, widths, and depths of the gaps. We use Monte Carlo radiative transfer codes to produce synthetic observations of model protostars that can be fit to our combined millimeter visibility and broadband SED dataset of GY 91. This modeling procedure is described in further detail in \citet{Sheehan2014} and Sheehan \& Eisner (submitted), but we give a brief overview here. 

Our model includes a flared protoplanetary disk with a physically motivated surface density profile \citep[e.g.][]{LyndenBell1974} surrounded by a rotating collapsing envelope \citep[e.g.][]{Ulrich1976},
\begin{equation}
\Sigma = \Sigma_0 \, \left(\frac{R}{r_c}\right)^{-\gamma} \, \exp\left[-\left(\frac{R}{r_c}\right)^{2-\gamma}\right],
\end{equation}
\begin{equation}
%\rho_{disk} = \rho_0 \left(\frac{R}{R_0}\right)^{-\alpha} \, \exp\left(-\frac{1}{2}\left[\frac{z}{h(R)}\right]^2\right),
\rho_{disk} =\frac{\Sigma}{\sqrt{2\pi}\,h} \, \exp\left(-\frac{1}{2}\left[\frac{z}{h}\right]^2\right),
\end{equation}
\begin{equation}
h = h_0 \left(\frac{R}{1 \text{ AU}}\right)^{\beta},
\end{equation}
\begin{equation}
\footnotesize
\rho_{env} = \frac{\dot{M}}{4\pi}\left(G M_* r^3\right)^{-\frac{1}{2}} \left(1+\frac{\mu}{\mu_0} \right)^{-\frac{1}{2}} \left(\frac{\mu}{\mu_0}+2\mu_0^2\frac{R_c}{r}\right)^{-1}.
\end{equation}
In Equations 1, 2 \& 3, $R$ and $z$ are in cylindrical coordinates, while in Equation 4, $\mu = \cos\,\theta$ and $r$ and $\theta$ are in spherical coordinates. 

In this model the disk mass, $M_{disk}$, inner and outer radii, $R_{in}$ \& $R_{disk}$, surface density power-law exponent, $\gamma$, scale height power-law exponent, $\beta$, and scale height at 1 AU, $h_0$, are left as free parameters. We also leave the envelope mass, $M_{env}$, and radius, $R_{env}$, as free parameters, and give the envelope an outflow cavity described by $f_{cav}$, the fraction by which the density is reduced in the cavity, and $\xi$, which relates to the cavity opening angle. We supply the density structure with opacities described in \citet{Sheehan2014}, leaving the maximum dust grain size, $a_{max}$, and grain size distribution power-law exponent, $p$, as free parameters. 

We model the dark lanes as gaps in the surface density profile, which are described by the radius at the center of the gap ($R_{gap,i}$), width ($w_{gap,i}$), and depth ($\delta_{gap,i}$). The depth of the gap is a multiplicative factor that represents the amount by which the surface density is reduced in the gap. $\delta = 0$ corresponds to a complete absence of material in the gap. 

%Because full radiative transfer modeling is challenging, we make initial estimates of disk properties and the gap widths and depths by fitting a simple geometric model to the $870$ $\mu$m and 3 mm visibilities (see Figure 3) using the MCMC code \texttt{emcee} \citep{ForemanMackey2013}. The parameters found from this simple geometrical fit are then used as initial guesses to generate a detailed radiative transfer model that simultaneously matches both visibility datasets and the broadband SED. The purpose of the radiative transfer modeling is, in particular, to determine whether an envelope component is needed to explain the observations. As such, we fix the gap properties for the radiative transfer modeling based on the results of the geometric fit. Furthermore, as radiative transfer models are computationally expensive and modeling heteroscedastic datasets is challenging, we have opted to find a model that matches the datasets well by-hand. We use the Monte Carlo radiative transfer codes RADMC-3D \citep{Dullemond2012} and Hyperion \citep{Robitaille2011} to calculate the temperature throughout the density structure described above, and then to produce synthetic millimeter visibilities and broadband SEDs.}

Because full radiative transfer modeling is very computationally intensive, we make initial estimates of disk properties and the gap widths and depths by fitting a simple geometric model to the $870$ $\mu$m and 3 mm visibilities (see Figure 3) using the MCMC code \texttt{emcee} \citep{ForemanMackey2013}. The parameters found from this simple geometrical fit are then used as initial guesses to generate a detailed radiative transfer model that simultaneously matches both visibility datasets and the broadband SED. 

The purpose of the radiative transfer modeling is, in particular, to determine whether an envelope component is needed to explain the observations. We use the Monte Carlo radiative transfer codes RADMC-3D \citep{Dullemond2012} and Hyperion \citep{Robitaille2011} to calculate the temperature throughout the density structure described above, and then to produce synthetic millimeter visibilities and broadband SEDs. We fit these synthetic observations simultaneously to all three (870 $\mu$m visibilities, 3 mm visibilities, and broadband SED) of our datasets, again using the \texttt{emcee} code.

The MCMC walkers are allowed to move through parameter space for an extended burn-in period. We consider the fit to have converged when the walkers have reached a steady state, with the best fit value for each parameter changing minimally over a large number of steps. We then calculate the best-fit value for each parameter as the median position of the walkers after discarding the burn-in steps. We estimate the uncertainty on these parameter values as the range around the median containing 68\% of the walkers.

\begin{figure*}
\centering
\figurenum{4}
\includegraphics[width=7in]{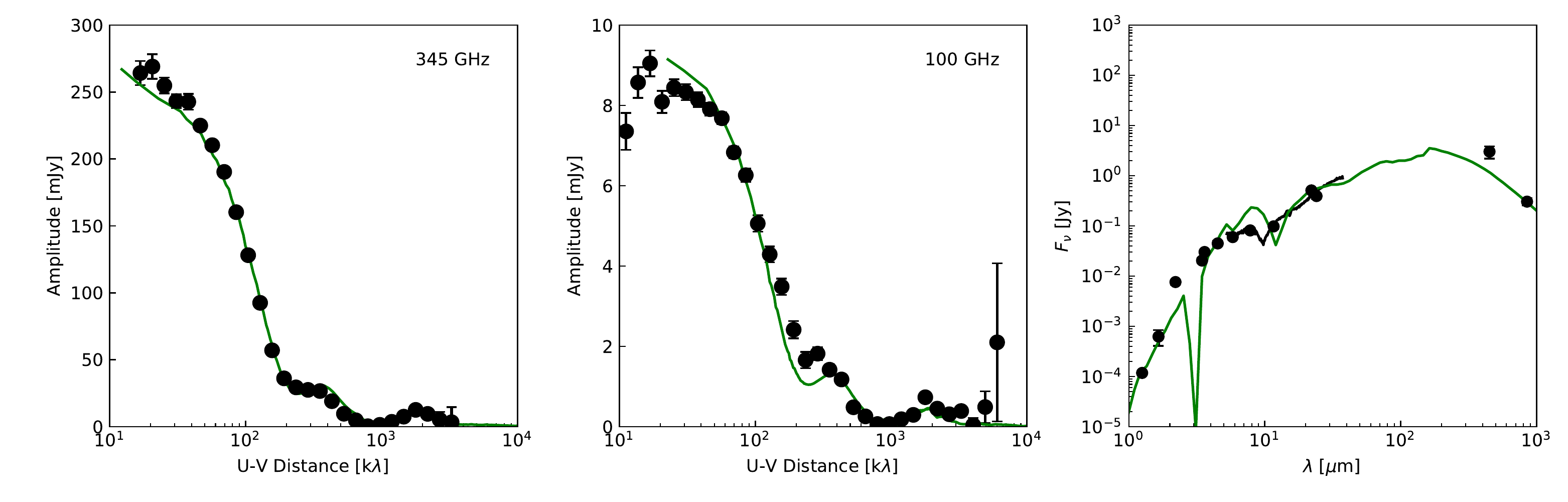}
\caption{The gapped disk+envelope model for GY 91 compared with the data. We show the one dimensional, azimuthally averaged, 870 $\mu$m visibility amplitudes on the left, the 3 mm visibilities in the center, and the SED on the right.}
\label{fig:disk_ulrich}
\end{figure*}

We compare the gapped disk+envelope model to the observed visibilities and SED in Figure \ref{fig:disk_ulrich} and list the model parameters in Table \ref{table:best_fits}. The images for our gapped disk+envelope model look almost identical to those shown in Figure \ref{fig:geometrical_model}, although the residuals between data and model are higher, not surprising since we are fitting the visibilities and SED simultaneously here.

This model can simultaneously reproduce the 870 $\mu$m visibilities, 3 mm visibilities, and broadband SED for GY 91. The GY 91 disk appears to be embedded in an envelope with $M_{env} = 0.12 \, M_{disk}$. 
%The envelope mass and radius are degenerate due to our limited constraint on large scale structure by the visibilities. However, we can reasonably constrain that an envelope envelope with $M_{env} \gtrsim 0.032$ M$_{\odot} \text{ } (\gtrsim 0.25 M_{disk})$ and $R_{env} \gtrsim 1000$ AU are needed.} 
Gaps are found at radii of $\sim$7 AU, $\sim$40 AU, and $\sim$69 AU, with widths of $\sim$7 AU, $\sim$30 AU, and $\sim$10 AU. The gap depths for the inner and outer gaps are not well constrained because they are not resolved well by our observations. The middle gap appears to be wide and somewhat shallow, although with higher resolution it is possible that it will break up into multiple gaps.

%\floattable
\input{table2.tex}

\section{Discussion \& Conclusion}

%\begin{figure*}
%\centering
%\figurenum{5}
%\includegraphics[width=7in]{figure3.pdf}
%\caption{A comparison of the best fitting disk model to the data. In the top row we show fits and models for the 345 GHz data and in the bottom row we show the fits and models for the 100 GHz data. We show the images in the left column with red dashed ellipses showing the locations of the gaps. The center column shows the best fit model images at each wavelength, convolved with the clean beam from the relevant dataset. Finally, the right column shows residuals produced by subtracting the best fit model from the data in the visibility plane and Fourier transforming to produce an image. The peak 345 GHz residual is 6.5 $\sigma$, and the peak 100 GHz residual is 7.9 $\sigma$.}
%\label{fig:models}
%\end{figure*}

GY 91 appears to be part of a growing population of protoplanetary disks that have ring-like features in their millimeter emission profiles. The 870 $\mu$m image resembles the disks of HL Tau, AA Tau, TW Hya, HD 163296, HD 169142, AS 209, and Elias 24, all of which have several gaps visible in their millimeter emission profiles \citep{Brogan2015,Andrews2016,Isella2016,Loomis2017,Fedele2017,Fedele2017b,Dipierro2018}. Closer inspection of these systems, however, reveals differences in the appearance of the features in each disk. The bright and dark rings seen in TW Hya are narrow (sizes $<2$ AU) and shallow \citep{Andrews2016}. Only the innermost gap, at 2 AU, has a significant depth. The gaps found in AA Tau, HD 163296, and HD 169142, on the other hand, are all very wide and deep, with widths of 22 -- 55 AU \citep{Isella2016,Loomis2017,Fedele2017}. The gaps found in HL Tau and AS 209 appear to be deep, with moderate widths of $\sim5-30$ AU \citep[e.g.][]{Brogan2015,Zhang2015}.

The innermost and outermost gaps we find in GY 91's disk appear to be quantitatively the most similar to the HL Tau gaps as they are somewhat narrow, with widths of $\sim$7 AU and $\sim$10 AU, while the middle gap appears to be large like the gaps found in AA Tau, HD 163296, and HD 169142. 

\subsection{Planets Carving Gaps?}

Although there are a number of potential origins of these features, the most exciting possibility is, perhaps, that these gaps are carved by proto-planets embedded in the disk. \citet{Dong2015} found that the gaps in the HL Tau disk could be sculpted by planets with masses as small as a Saturn-mass. \citet{Isella2016} found similar results for HD 163296, although the gaps are much larger in that disk.

We can estimate the masses of planets that are needed to produce the gaps we see in GY 91's disk. Simulations suggest that planets should open gaps whose widths are a few times larger than the Hill radius of the planet,
\begin{equation}
W \approx 8 \times R_p \, \left(\frac{M_p}{M_*}\right)^{1/3}
\end{equation}
\citep{Rosotti2016}. Although the protostellar mass of GY 91 is not constrained well, it is thought to be a M4 protostar with a temperature of 3300 K \citep{Doppmann2005}, which evolutionary models predict should have a mass of $\sim0.25$ M$_{\odot}$ at $\sim0.5$ Myr \citep{Baraffe2015}. Using these assumptions, we estimate that planets of masses $\sim0.2$ M$_J$, $\sim$0.2 M$_J$, and $\sim$0.002 M$_J$ are needed to produce the observed gaps. 

The mass estimated for the outermost planet highlights the limitations of these simple estimates, as it seems unlikely that an Earth-mass planet is opening such a gap. Recent studies have suggested that for low-mass planets, the gap width may be a constant multiple of the scale height and therefore independent of planet mass \citep[e.g.][]{Duffell2013,Dong2017}. Further studies suggest that the mass of a gap-opening planet is best constrained by measurements of the gap width and depth in the gas distribution, combined with a measurement of disk viscosity \citep{Fung2014,Kanagawa2015,Dong2017}. Without knowledge of the gas distribution, however, we cannot place stronger constraints on the potential planet masses. It also should be noted that a single planet can open multiple gaps in a disk \citep{Bae2017}.

\subsection{Other Causes of Dark Lanes}

Planets aren't the only possible cause of these features. One alternative that should be common in protoplanetary disk is the variation in dust opacities and collisional fragmentation/coagulation properties that is expected to occur at snow lines. As dust grains radially drift inwards due to the loss of angular momentum from a headwind of sub-Keplerian gas \citep{Weidenschilling1977b}, they will cross a series of snow lines for various volatiles. When they cross a snow line, that volatile sublimates back into the gas phase. As the sublimated gas radially diffuses, it can re-condense onto particles outside of the snow line. The icy particles outside of the snow line can efficiently grow to decimeter or larger sizes, while solids inside the snow line tend to fragment \citep{Cuzzi2004,Ros2013,Banzatti2015}. The change in the optical properties of dust grains across the snow line could cause features like those seen in HL Tau or GY 91 \citep[e.g.][]{Zhang2015}. 

Alternatively, the ``sintering" of dust grains just below the sublimation temperature produces brittle grains that fragment more readily and therefore grow to smaller sizes just outside the snow line. Because the sintered grains have smaller sizes, they undergo slower radial drift, causing pile ups near snow lines. This process could also produce features similar to those seen in HL Tau or GY 91 \citep{Okuzumi2016}.

\begin{figure*}[t]
\centering
\figurenum{5}
\includegraphics[width=6in]{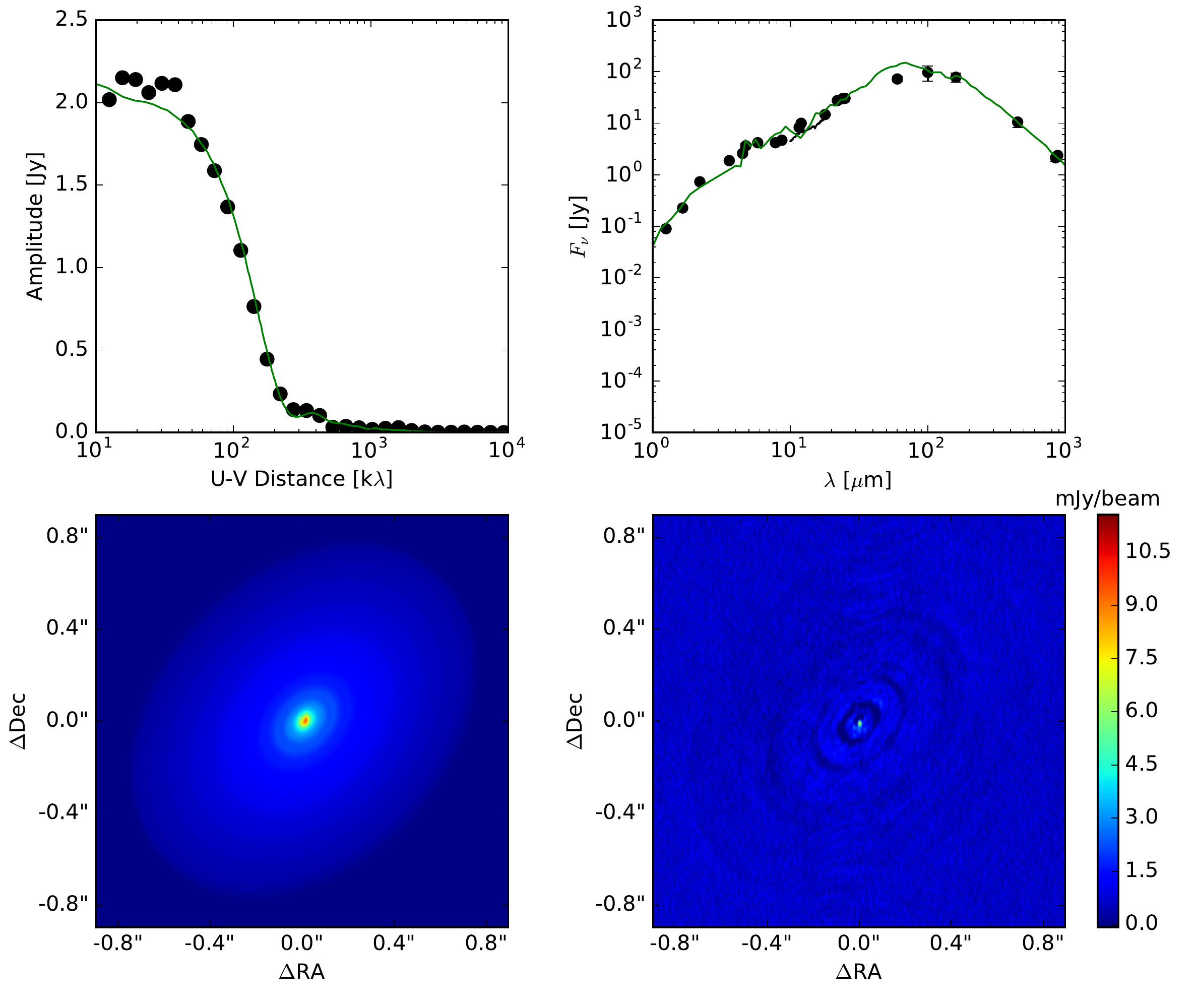}
\caption{The disk+envelope model for HL Tau compared with the data. In the first row we show the one dimensional, azimuthally averaged, 870 $\mu$m visibility amplitudes on the left and the SED on the right, with the model as a green curve in both. The second row shows the 345 GHz model and residual images. We did not include gaps in this model, which is why they can be seen in the residual map. The model has a disk with a mass of 0.2 M$_{\odot}$ a radius of 120 AU, a surface density power law exponent of $\gamma = 1.7$, and an inclination of 44$^{\circ}$. The envelope has a mass of $0.04$ M$_{\odot}$ and a radius of 1800 AU.}
\label{fig:hltau}
\end{figure*}

We compare the midplane disk temperature inferred for GY 91 with the temperatures of snow lines of common volatiles \citep[e.g.][]{Zhang2015}. The outermost gap does roughly match the freeze out region of $N_2$. However no obvious counterparts are seen for the inner two gaps.

Zonal flows produced by magneto-rotational instability driven turbulence \citep{Johansen2009} have also been shown to produce axisymmetric pressure bumps that can trap large dust grains and may produce gap-like features in millimeter images \citep{Pinilla2012,Dittrich2013,Simon2014}. In this case, the pressure bumps are created by large scale variations in the magnetically driven turbulence that produce variations in the mass accretion rate that in turn causes material to pile up. This effect can also be seen at the outer edge of magnetic dead zones, where there is strong radial variation in the mass accretion rate. These flows can produce gap-like features in disks \citep{Pinilla2012,Flock2015}.

\subsection{Comparison with HL Tau}

If planets are indeed carving gaps in GY 91's disk, the masses of those planets would place strong constraints on the timescales for planet formation in disks. As a Class I protostar, GY 91 likely has an age of $\sim0.5$ Myr \citep{Evans2009}, so planets must grow to masses of $\sim0.2$ M$_J$ on these short timescales. 

Similar constraints have been placed on the timescale for planet formation by the gaps in other young disks. AS 209 and Elias 24 are also in $\rho$ Ophiuchus, which has been estimated to be quite young \citep[$0.5-1$ Myr; e.g.][]{Natta2006}. Both are, however, found to be Class II sources \citep[e.g.][]{Barsony2005,Andrews2009}, and therefore are likely to be older than GY 91. HL Tau's disk may be even younger than AS 209 and Elias 24, as it is also possibly still embedded, but comparable to or perhaps older than GY 91 \citep[e.g.][]{Robitaille2007}. 

To the best of our knowledge, however, a detailed radiative transfer modeling fit to the combined HL Tau millimeter visibilities and SED has not been done since the ALMA Science Verification data was acquired. We use the disk+envelope modeling procedure described above to fit a disk+envelope model to the HL Tau ALMA millimeter visibilities and SED. For simplicity, though, we ignore the gaps and consider only a smooth density distribution. Furthermore, as it is possible that we are resolving out large scale envelope structure with our visibility datasets, we have limited the HL Tau visibilities to a minimum baseline of 16 k$\lambda$ to match our GY 91 dataset and ensure a fair comparison of the two sources. The best fit model is shown in Figure \ref{fig:hltau}. 

\begin{figure}[t]
\centering
\figurenum{6}
\includegraphics[width=3.25in]{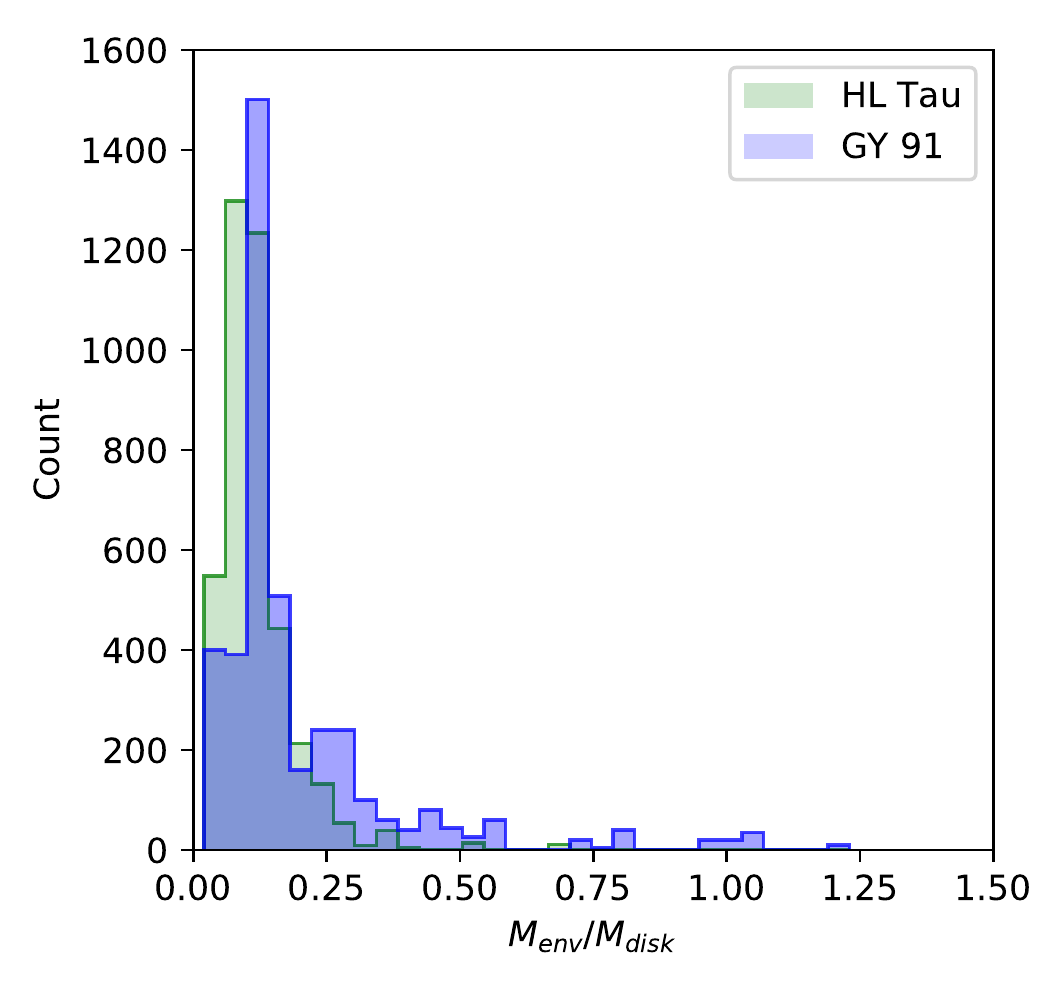}
\caption{Histograms of the positions of MCMC walkers in $M_{env}/M_{disk}$ over the last 10 steps of the MCMC fit, indicating the allowed range $M_{env}/M_{disk}$ for each source.}
\label{fig:menv_mdisk}
\end{figure}

Our model for HL Tau has $M_{env} = 0.106 \, M_{disk}$, similar to what we find for GY 91 ($M_{env} = 0.121 \, M_{disk}$). This is in agreement with the classification of HL Tau as a ``flat spectrum" object, indicating it is likely in transition from the Class I to the Class II stage. The range of allowed values for $M_{env}/M_{disk}$ are shown in \ref{fig:menv_mdisk}. Although the distributions are similar, suggesting that both sources are at a similar stage of their evolution, our modeling finds a slightly larger value of $M_{env}/M_{disk} = 0.121$ for GY 91. This may indicate that a larger fraction of the HL Tau envelope has been depleted onto the disk or central protostar, although not by much. 

If we assume that $M_{env}/M_{disk}$ is an evolutionary indicator \citep[e.g.][]{Crapsi2008}, this suggests that GY 91 is of similar age to, or possibly younger than HL Tau. 
%It is possible that we are underestimating the envelope masses of both sources, as spatial filtering likely resolves out emission from scales $\gtrsim1000$ AU, however we still expect the ratio of the envelope mass within 1000 AU to disk mass to be an indicator of evolution}. 
The identification of multiple young disks with gaps is a clear indication that planet formation, or at the very least processes that contribute to planet formation, are happening in the very early stages of stellar formation. If GY 91 is even younger than HL Tau, then measurements of the masses of planets embedded in it's disk could place even stronger constraints on the timescales of planet formation than planets in the HL Tau disk.

\vspace{10pt}

Regardless of whether these dark lanes are formed by planets, zonal flows, or chemical variations produced by radial drift, the presence of these features is likely an indication that planet formation is well underway at early times. Both zonal flows and chemical effects have been suggested to enhance the growth of particles in disks \citep[e.g.][]{Simon2014,Ros2013}, and may be key elements in how planets are formed. Further high resolution studies of these young disks are crucial for understanding the early stages of planet formation.

\acknowledgements

The authors would like to thank the anonymous referee, whose comments helped to improve this work. This work was supported by NSF AAG grant 1311910. This paper makes use of the following ALMA data: ADS/JAO.ALMA\#2015.1.00761.S. ALMA is a partnership of ESO (representing its member states), NSF (USA) and NINS (Japan), together with NRC (Canada), NSC and ASIAA (Taiwan), and KASI (Republic of Korea), in cooperation with the Republic of Chile. The Joint ALMA Observatory is operated by ESO, AUI/NRAO and NAOJ The National Radio Astronomy Observatory is a facility of the National Science Foundation operated under cooperative agreement by Associated Universities, Inc.

\software{CASA \citep{McMullin2007}, RADMC-3D \citep{Dullemond2012}, Hyperion \citep{Robitaille2011}, emcee \citep{ForemanMackey2013}}

%\clearpage

\bibliography{ms.bib}

\end{document}

%% file: table1.tex
\begin{deluxetable}{lcccc}
\tablenum{1}
\tablecaption{Log of ALMA Observations}
\label{table:alma_obs}
\tablehead{\colhead{Observation Date} & \colhead{ALMA Band} & \colhead{Baselines} & \colhead{Total Integration Time} & \colhead{Calibrators} \\
\colhead{(UT)} & \colhead{} & \colhead{(m)} & \colhead{(s)} & \colhead{(Flux, Bandpass, Gain)}}
\startdata
Oct. 31 2015 & 3 & 84$\,$-$\,$16,200 & 169 & 1517-2422, 1625-2527 \\
Nov. 26 2015 & 3 & 68$\,$-$\,$14,300 & 169 & 1517-2422, 1625-2527 \\
Apr. 17 2016 & 3 & 15$\,$-$\,$600 & 58 & 1733-1304, 1427-4206, 1625-2527 \\
May 19 2016 & 7 & 15$\,$-$\,$640 & 30 & J1517-2422, J1625-2527 \\
Sep. 11, 2016 & 7 & 15$\,$-$\,$3140 & 60 & J1517-2422, J1625-2527 \\
\enddata
\end{deluxetable}

%% file: table2.tex
\begin{deluxetable}{lccccccccccccccccccccc}
\tablecaption{Gapped Disk+Envelope Model Parameters}
\tablenum{2}
\label{table:best_fits}
\tablehead{\colhead{Parameters} & \colhead{Values}}
\startdata
$L_{star}$ [L$_{\odot}$] & $0.11 \pm 0.00$ \\
$M_{disk}$ [$M_{\odot}$] & $ 0.17 \pm  0.00$ \\
$R_{in}$ [AU] & $0.1 \pm 0.0$ \\
$R_{disk}$ [AU] & $   75 \pm     1$ \\
$h_0$ [AU] & $0.29 \pm 0.01$ \\
$\gamma$ & $0.01 \pm 0.04$ \\
$\beta$ & $0.30 \pm 0.01$ \\
$M_{env}$ [$M_{\odot}$] & $0.020 \pm 0.008$ \\
$R_{env}$ [AU] & $1269 \pm  328$ \\
$f_{cav}$ & $0.56 \pm 0.04$ \\
$\xi$ & $1.31 \pm 0.24$ \\
$i$ [$^{\circ}$] & $41 \pm  1$ \\
p.a. [$^{\circ}$] & $111 \pm   1$ \\
$a_{max}$ [$\mu$m] & $82597 \pm  6318$ \\
$R_{gap,1}$ [AU] & $ 10.0 \pm   0.2$ \\
$w_{gap,1}$ [AU] & $ 7.0 \pm  0.4$ \\
$\delta_{gap,1}$ & $0.01 \pm 0.02$ \\
$R_{gap,2}$ [AU] & $ 40.9 \pm   0.4$ \\
$w_{gap,2}$ [AU] & $30.0 \pm  6.7$ \\
$\delta_{gap,2}$ & $0.16 \pm 0.07$ \\
$R_{gap,3}$ [AU] & $ 69.4 \pm   0.6$ \\
$w_{gap,3}$ [AU] & $10.0 \pm  0.8$ \\
$\delta_{gap,3}$ & $0.01 \pm 0.02$ \\
$p$ & $3.50$
\enddata
\end{deluxetable}